\documentclass[runningheads]{llncs}
\input{psfig.sty}
\usepackage{amssymb}

\usepackage[utf8]{inputenc}
\usepackage{amsmath}
\usepackage{amsfonts}
\usepackage{amssymb}
\usepackage{graphicx}

\newcommand{\bea}{\begin{eqnarray*}}
\newcommand{\eea}{\end{eqnarray*}}

\newcommand{\teta}{\theta}

\newtheorem{lemmma}{Lemma}

\input{Qcircuit.tex}

\begin{document}

\mainmatter
\title{From reversible computation to quantum computation by Lagrange interpolation}
\titlerunning{Lagrange interpolation}
\author{Alexis De Vos$^1$ and Stijn De Baerdemacker$^2$} 
\authorrunning{A. De Vos and S. De Baerdemacker}
\institute{$^1$ Cmst, Imec v.z.w., \\
           vakgroep elektronica en informatiesystemen, \\ 
           Universiteit Gent, B - 9000 Gent, Belgium \\
           {\tt alex@elis.ugent.be}  \\
           $^2$ Ghent Quantum Chemistry Group, \\
           vakgroep anorganische en fysische chemie, \\
           Universiteit Gent, B - 9000 Gent, Belgium \\
           {\tt Stijn.DeBaerdemacker@ugent.be} } 

\maketitle

\begin{abstract}
Classical reversible circuits, acting on $w$~bits,
are represented by permutation matrices of size $2^w \times 2^w$.
Those matrices form the group P($2^w$), 
isomorphic to the symmetric group {\bf S}$_{2^w}$. 
The permutation group P($n$), isomorphic to {\bf S}$_n$,
contains cycles with length~$p$, ranging from~1 to $L(n)$, 
where $L(n)$ is the so-called Landau function.
By Lagrange interpolation between the $p$~matrices of the cycle,
we step from a finite cyclic group of order~$p$ to 
a 1-dimensional Lie group, subgroup of the unitary group U($n$). 
As U($2^w$) is the group of all possible quantum circuits, acting on $w$~qubits,
such interpolation is a natural way to step 
from classical computation to quantum computation. 
\end{abstract}

\section{Introduction}

Too often conventional (every-day) computers and (futuristic) quantum computers
are considered as two separate worlds, far from each other.
Conventional computers act on classical bits, say `pure' zeroes and ones,
by means of Boolean logic gates, such as {\tt AND}~gates and {\tt OR}~gates \cite{boek}.
The operations performed by these gates usually are described by truth tables.
Quantum computers act on qubits, say complex vectors,
by means of quantum gates, such as {\tt ROTATOR} gates and {\tt T}~gates \cite{nielsen}. 
The operations performed by these gates usually are described by unitary matrices.

Because 
the world of classical computation and 
the world of quantum   computation are based on such different science models,
it is difficult to see the relationship (both analogies and differences) 
between these two computation paradigms.
In order to remedy this problem, in recent years, 
{\tt NEGATOR} gates and {\tt PHASOR} gates have been developed as basic building-blocks
of quantum circuits \cite{bremen} \cite{joris} \cite{acm}.
In the present paper, we bridge the gap between the two sciences
by deriving the {\tt NEGATOR} gates in an alternative way.
The common tool we have chosen for describing both
reversible computation and quantum computation is the matrix representation.
Classical reversible circuits \cite{boek}, acting on $w$   bits, are represented
by permutation matrices of size $2^w \times 2^w$, whereas 
quantum              circuits \cite{nielsen}, acting on $w$ qubits, are represented
by unitary     matrices of size $2^w \times 2^w$.
Invertible matrices form a group under the operation of matrix multiplication.
The matrix group consisting of permutation matrices
is a subgroup of the group of unitary matrices.
In the present paper, we show how to enlarge the subgroup to its supergroup,
in other words: how a classical computer can be upgraded to a quantum computer.
It is surprising that a mathematical tool for this tour-de-force
is the good-old polynomial interpolation formula of Lagrange.
By Lagrange interpolation between two (or more) permutation matrices 
we indeed obtain an infinity of unitary matrices.  

\section{The Lagrange interpolation}

We consider the $n \times n$ unitary matrix $q$.
The only restriction is the finiteness of its order.
In other words:
we assume that $q^{\, p}$ equals the $n \times n$ unit matrix~$u$
and that none of the matrices $q^{\, s}$ with $0 < s < p$  equals~$u$.
Thus the set $\{ q, q^2, q^3, ..., q^{p-1}, q^p \}$ (with $q^p=q^0=u$)
constitutes a finite matrix group 
isomorphic to the cyclic group {\bf Z}$_p$ of order~$p$.

\begin{figure}[!htb]
\begin{center}
\includegraphics{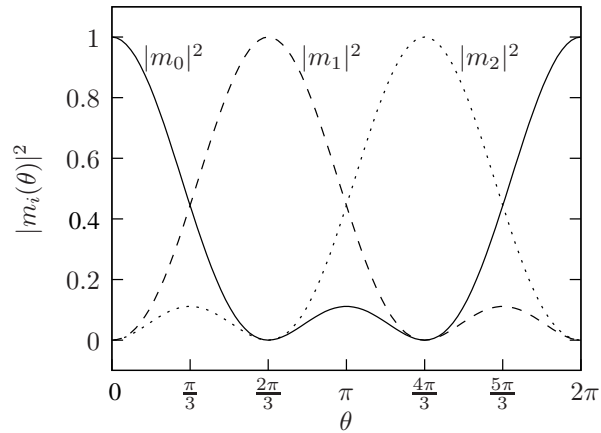}
\caption{The moduli squared of the three coefficients $m_j(\teta)$ in $\sum_{j=0}^2 m_j(\teta) q^j$,
         the Lagrange interpolation
         between the three unitary matrices $q^0$, $q^1$, and $q^2$. }
\label{fig:lagrange}
\end{center}
\end{figure}

We construct a 1-dimensional Lie group which constitutes a smooth interpolation
between the $p$ matrices $q^j$. For this purpose, we are
inspired by the interpolation formula of Lagrange:
if, for the values $x=x_j$, a function $y(x)$ evaluates to~$y_j$,
then the polynomial
\[
p(x) = \sum_j \ \frac{\prod_{k \neq j} (x-x_k)}{\prod_{k \neq j} (x_j-x_k)}\ y_j
\]
automatically satisfies $p(x_j)=y_j$, for all~$x_j$.
The function $p(x)$ connects the points $(x_j, y_j)$ in an analytic and smooth way. 
By analogy, we construct the matrix
\begin{equation}
m(\teta) = \sum_j \ \frac{\prod_{k \neq j} (e^{i\teta}-\omega^k)}{\prod_{k \neq j} (\omega^j-\omega^k)}\ q^j \ ,
\label{1}
\end{equation}
where the constant $\omega$ is the primitive $p$~th root of unity, i.e.\ $e^{i2\pi/p}$. 
Here, indices\footnote{Also in the rest of the paper, 
                       each time the limits of a summing index are not specified, 
                       they equal~0 and $p-1$.} 
($j$ and $k$) run from~0 to $p-1$. 
The reader can easily verify that indeed we automatically have 
$m(j\, 2\pi/p) = q^j$. In particular we have $m(0)=q^0=u$.
Expression (\ref{1}) constitutes a generalization of the case $p=2$, 
discussed in Appendix~A of \cite{acm}. 
Indeed, for $p=2$, eqn (\ref{1}) recovers the expression found in \cite{acm}:
\begin{equation}
m(\teta) = \frac{1+e^{i\teta}}{2}\ q^0 + \frac{1-e^{i\teta}}{2}\ q^1 \ .
\label{p2}
\end{equation}
Rewriting (\ref{1}) as
\begin{equation}
m(\teta) = \sum_j m_j(\teta)\, q^j \ ,
\label{m}
\end{equation}
the coefficients $m_j(\teta)$ 
(known as the Lagrange basis polynomials or Lagrange fundamental polynomials~\cite{davis}\,) 
have the property that
$m_j(k\, 2\pi/p)$ equals~1 for $k=j$ but equals~0 otherwise.
This is illustrated for $p=3$ in Figure~\ref{fig:lagrange},
where we see the modulus squared of the three coefficients
\bea
m_0(\teta) & = &  \frac{e^{i\teta} - \omega  }{1        - \omega  } \  \frac{e^{i\teta}-\omega^2}{1        - \omega^2} \\[0.5mm]
m_1(\teta) & = &  \frac{e^{i\teta} - 1       }{\omega   - 1       } \  \frac{e^{i\teta}-\omega^2}{\omega   - \omega^2} \\[0.5mm]
m_2(\teta) & = &  \frac{e^{i\teta} - 1       }{\omega^2 - 1       } \  \frac{e^{i\teta}-\omega  }{\omega^2 - \omega  } \ ,
\eea
where $\omega$ is the primitive cubic root of unity, i.e.\ $e^{i\, 2\pi/3} = - 1/2 + i\,\sqrt{3}/2$.

There exist various different expressions equivalent to (\ref{1}):
\begin{eqnarray}
m(\teta) & = & \frac{1}{p}\ \sum_j \ \omega^j\, q^j\, \prod_{k \neq j} (e^{i\teta}-\omega^k)       \label{2} \\[0.5mm]
m(\teta) & = & \frac{1}{p}\ (e^{ip\teta} - 1)\ \sum_j \ \frac{\omega^j}{e^{i\teta}-\omega^j}\ q^j  \label{3} \\[0.5mm]
m(\teta) & = & \frac{1}{p}\ \sum_j q^j \sum_r \ \omega^{-rj} e^{ir\teta} \ .                       \label{4}
\end{eqnarray}
Appendix~A describes how they are deduced from (\ref{1}).
Eqn~(\ref{2}) is more compact than eqn~(\ref{1}), 
as it contains only one instead of two products.
Expression (\ref{3}) corresponds with the so-called
`first barycentric form' of the Lagrange interpolation formula.
It has the advantage that it contains no product at all.
However, it has a disadvantage with respect to~(\ref{2}):
in eqn~(\ref{2}) we clearly see that $m$ is a polynomial (of degree $p-1$) in the variable $e^{i\teta}$;
in eqn~(\ref{3}) this fact is hidden.
The expression (\ref{4}) constitutes a good compromise between (\ref{2}) and (\ref{3}):
we still recognize the polynomial nature of the expression,
while the sum of products in (\ref{2}) is simplified to a sum of sums in (\ref{4}).

In Appendix~B, we prove that $m(\teta)$ has the property
\[
m(\teta_1)m(\teta_2) = m(\teta_1+\teta_2)\ ,
\]
such that the multiplication of two $m(\teta)$ matrices yields a third $m(\teta)$ matrix. 
In particular we have $m(\teta)m(-\teta)=m(0)=u$,
such that $m(\teta)^{-1} = m(-\teta)$. 
Hence, all conditions are fulfilled to say that the set $m(\teta)$, 
together with the operation of `ordinary matrix multiplication', 
forms a group. 
It is a continuous group: a~1-dimensional Lie group. 
In Appendix~C we demonstrate that the matrices $m(\teta)$ are unitary.
We thus may conclude that the group is
isomorphic to the unitary group~U(1).

Any 1-dimensional Lie group $m(\teta)$ has a generator:
\[
g = \frac{1}{i}\ \lim_{\teta \rightarrow 0} \frac{dm}{d\teta} \ .
\]
With the help of (\ref{3}) we find:
\bea
g & = & \sum_{j \neq 0}\ \frac{\omega^j}{1-\omega^j}\ q^ j + 
        \lim\, [\ \frac{e^{ip\teta}}{e^{i\teta}-1} - \frac{(e^{ip\teta}-1)e^{i\teta}}{p\, (e^{i\teta}-1)^2}\ ]\ q^0 \nonumber\\
  & = & \frac{p-1}{2}\ q^0 + \sum_{j \neq 0}\ \frac{\omega^j}{1-\omega^j}\ q^ j \ .
\eea
With the help of (\ref{4}) we find an alternative expression for this $n \times n$ matrix:
\[
g = \frac{1}{p}\ \sum_j\ q^j\ \sum_r\ r\, \omega^{-rj} \ .
\]
For the case $p=2$, both expressions simplify to $g=(q^0-q^1)/2$,
a result that can also be retrieved directly from (\ref{p2}).

\section{Examples}

The $n \times n$ unitary matrices form the unitary group U($n$).
Within this infinite group figures the finite subgroup P($n$) of 
all $n \times n$ permutation matrices. 
We choose as matrix $q$ of the previous section,
one of these $n!$ permutation matrices. Such matrix generates
a finite set of matrices $\{ q, q^2, ..., q^{p}\}$.
This set is a subgroup of P($n$), isomorphic to the cyclic group {\bf Z}$_p$ of order~$p$.
The value of $p$ depends on the particular choice of the permutation matrix~$q$.
The minimum value is~1 (for the trivial choice $q=u$);
the maximum value is $L(n)$, where $L$ denotes the Landau function \cite{landau}.

The group P($n$) is isomorphic to the symmetric group {\bf S}$_n$.
The cycle graph of a finite group depicts its cyclic subgroups.
By convention, only the primitive or maximal cycles 
(i.e.\ those cycles that are not subsets of another cycle) are shown.

As a first example, we investigate P(2), i.e.\ the group of all $2 \times 2$ permutation matrices.
It contains two elements, forming a single 2-cycle:
\begin{quote} 
      $\bullet$ the two matrices $q   = \left( \begin{array}{cc} 0 & 1 \\ 1 & 0 \end{array} \right)$ and 
                                 $q^2 = \left( \begin{array}{cc} 1 & 0 \\ 0 & 1 \end{array} \right) = q^0$\ ,
\end{quote} 
the matrix $q$ representing the {\tt NOT} gate.
After (\ref{p2}), the interpolation between $q^0$ and $q^1$ is
\[
m(\teta) = \frac{1}{2}\ \left( \begin{array}{cc} 1 + e^{i\teta} & \ 1 - e^{i\teta} \\ 1 - e^{i\teta} & \ 1 + e^{i\teta} \end{array} \right) \ ,
\]
called the {\tt NEGATOR} gate \cite{negator}. It is a quantum gate, generalization of the classical gates
$m(0)$ (i.e.\ the {\tt IDENTITY} gate) and $m(\pi)$ (i.e.\ the {\tt NOT} gate). 
We note that, for any value of the angle~$\teta$, both row sums and both column sums of $m(\teta)$
are equal to~1. The generator of $m(\teta)$ is
\[
g = \frac{1}{2}\ \left( \begin{array}{rr} 1 & \ -1 \\ -1 & 1 \end{array} \right)\ ,
\]
a matrix with all line sums equal to~0.

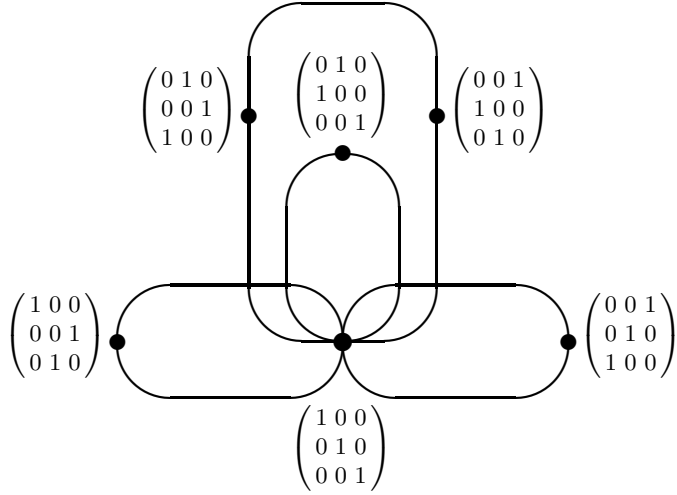
\begin{figure}[tb]
\vspace*{9mm}
\setlength{\unitlength}{0.50mm}
\begin{picture}(120,120)(-40,0)
\thicklines
\put( 80, 40){\circle*{5}}
\put( 66, 10){$\left(\begin{array}{ccc} 1 & 0 & 0 \\ 0 & 1 & 0 \\ 0 & 0 & 1 \end{array}\right)$}
\put( 80, 65){\oval(30,50)}
\put( 80, 90){\circle*{4}}
\put( 66,104){$\left(\begin{array}{ccc} 0 & 1 & 0 \\ 1 & 0 & 0 \\ 0 & 0 & 1 \end{array}\right)$}
\put( 50, 40){\oval(60,30)}
\put( 20, 40){\circle*{4}}
\put(-10, 40){$\left(\begin{array}{ccc} 1 & 0 & 0 \\ 0 & 0 & 1 \\ 0 & 1 & 0 \end{array}\right)$}
\put(110, 40){\oval(60,30)}
\put(140, 40){\circle*{4}}
\put(143, 40){$\left(\begin{array}{ccc} 0 & 0 & 1 \\ 0 & 1 & 0 \\ 1 & 0 & 0 \end{array}\right)$}
\put( 80, 85){\oval(50,90)}
\put(105,100){\circle*{4}}
\put( 55,100){\circle*{4}}
\put(108,100){$\left(\begin{array}{ccc} 0 & 0 & 1 \\ 1 & 0 & 0 \\ 0 & 1 & 0 \end{array}\right)$}
\put( 25,100){$\left(\begin{array}{ccc} 0 & 1 & 0 \\ 0 & 0 & 1 \\ 1 & 0 & 0 \end{array}\right)$}
\end{picture}
\caption{Cycle graph of the group P(3) of $3 \times 3$ permutation matrices.}
\label{graf1}
\end{figure}

As a second example, we take P(3), i.e.\ the group of all $3 \times 3$ permutation matrices.
It consists of $3! = 6$ elements, ordered in four maximal cycles:
\begin{itemize}
\item three 2-cycles:
      \begin{itemize}
      \item the two matrices $q   = \left( \begin{array}{ccc} 1 & 0 & 0 \\ 0 & 0 & 1 \\ 0 & 1 & 0 \end{array} \right)$ and 
                             $q^2 = \left( \begin{array}{ccc} 1 & 0 & 0 \\ 0 & 1 & 0 \\ 0 & 0 & 1 \end{array} \right) = q^0$ \vspace*{2mm}
      \item the two matrices $q   = \left( \begin{array}{ccc} 0 & 1 & 0 \\ 1 & 0 & 0 \\ 0 & 0 & 1 \end{array} \right)$ and
                             $q^2 = \left( \begin{array}{ccc} 1 & 0 & 0 \\ 0 & 1 & 0 \\ 0 & 0 & 1 \end{array} \right) = q^0$ \vspace*{2mm}
      \item the two matrices $q   = \left( \begin{array}{ccc} 0 & 0 & 1 \\ 0 & 1 & 0 \\ 1 & 0 & 0 \end{array} \right)$ and 
                             $q^2 = \left( \begin{array}{ccc} 1 & 0 & 0 \\ 0 & 1 & 0 \\ 0 & 0 & 1 \end{array} \right) = q^0$ \vspace*{1mm}
      \end{itemize}  
\item one 3-cycle, consisting of the three matrices
      \[
      q   = \left( \begin{array}{ccc} 0 & 1 & 0 \\ 0 & 0 & 1 \\ 1 & 0 & 0 \end{array} \right), \ \ 
      q^2 = \left( \begin{array}{ccc} 0 & 0 & 1 \\ 1 & 0 & 0 \\ 0 & 1 & 0 \end{array} \right), \mbox{ and } \ 
      q^3 = \left( \begin{array}{ccc} 1 & 0 & 0 \\ 0 & 1 & 0 \\ 0 & 0 & 1 \end{array} \right) = q^0 \ .
      \]
\end{itemize}
There exist no longer cycles, as $L(3)=3$.
Figure \ref{graf1} displays the graph.

We now examine the 3-cycle in detail. 
After (\ref{3}) and (\ref{4}), respectively, its unitary interpolation is  
\bea
m(\teta) & = & \frac{1}{3}\ (e^{i3\teta} - 1) \ \left( \ \frac{1}       {e^{i\teta}-1       }\ q^0 +
                                                         \frac{\omega}  {e^{i\teta}-\omega  }\ q^1 +
                                                         \frac{\omega^2}{e^{i\teta}-\omega^2}\ q^2 \ \right) \\[2mm]
         & = & \frac{1}{3}\ [\ (         e^{i2\teta} +          e^{i\teta} + 1)\ q^0 + 
                               (\omega   e^{i2\teta} + \omega^2 e^{i\teta} + 1)\ q^1 + 
                               (\omega^2 e^{i2\teta} + \omega   e^{i\teta }+ 1)\ q^2 \ ] \\[2mm]
         & = & \frac{1}{3}\  \left( \begin{array}{ccccc} 
                             x^2 + x + 1 && \omega x^2 + \omega^2 x + 1 && \omega^2 x^2 + \omega x + 1 \\ 
                             \omega^2 x^2 + \omega x + 1 && x^2 + x + 1 && \omega x^2 + \omega^2 x + 1 \\ 
                             \omega x^2 + \omega^2 x + 1 && \omega^2 x^2 + \omega x + 1 && x^2 + x + 1 \end{array} \right) \ ,
\eea
where $x$ is a short-hand notation for $e^{i\teta}$ and 
$\omega$ again is the primitive cubic root of unity.
As expected, we have $m(0)=q^0$, $m(2\pi/3)=q^1$, $m(4\pi/3)=q^2$, and $m(2\pi)=q^0$ again.

We notice that all six line sums (i.e.\ three row sums and three column sums) of $m(\teta)$ are equal to~1.
Therefore, we say that $m$ belongs to the subgroup XU(3) of U(3), described in detail in Appendix~B of \cite{negator}.
Here, XU($n$) is the subgroup of U($n$) consisting of all $n \times n$ unitary matrices with
all $2n$ line sums (i.e.\ $n$~row sums and $n$~column sums)
equal to~unity \cite{bremen} \cite{joris} \cite{acm}. 
This is no coincidence. In Appendix~D, we prove that, if $q$ is a unitary matrix from XU($n$),
then also the interpolation is a member of XU($n$).

For the generators of the four cycles we find:
\[
\hspace{-4mm}
\frac{1}{2}
\left( \begin{array}{ccc}   0 &  0 &  0 \\ 
                            0 &  1 & -1 \\ 
                            0 & -1 &  1 \end{array} \right) ,\,
\frac{1}{2}
\left( \begin{array}{ccc}   1 & -1 &  0 \\ 
                           -1 &  1 &  0 \\ 
                            0 &  0 &  0 \end{array} \right) ,\, 
\frac{1}{2}
\left( \begin{array}{ccc}   1 &  0 & -1 \\ 
                            0 &  0 &  0 \\ 
                           -1 &  0 &  1 \end{array} \right) ,\, \mbox{ and } \ 
\frac{1}{3}
\left( \begin{array}{ccc}           3 & \omega   - 1 & \omega^2 -1 \\ 
                          \omega^2 -1 &            3 & \omega  - 1 \\ 
                          \omega  - 1 & \omega^2 - 1 &           3 \end{array} \right) \ ,
\]
respectively.
Each has all line sums equal to~0.
Separately, each of these generators generates a 1-dimensional subgroup of XU(3);
together they generate the full 4-dimensional group XU(3), subgroup of the 9-dimensional group U(3). 

\section{Classical and quantum computing}

Because classical 
reversible circuits \cite{boek}    acting on $w$~bits,   are represented by matrices from P($2^w$) and
quantum    circuits \cite{nielsen} acting on $w$~qubits, are represented by matrices from U($2^w$),
we have special attention for the case $n=2^w$.

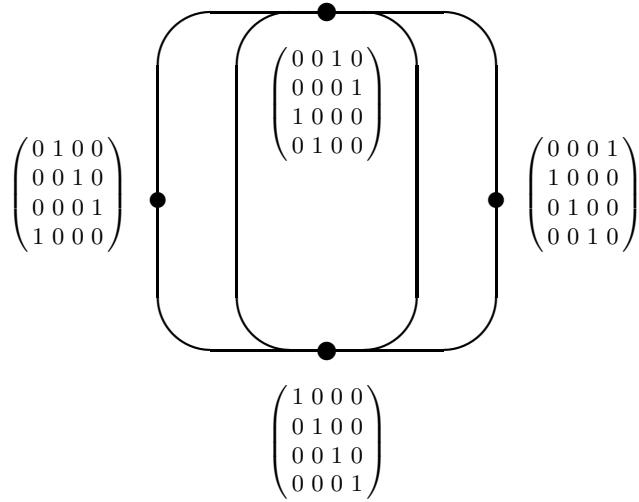
\begin{figure}[tb]
\vspace*{9mm}
\setlength{\unitlength}{0.50mm}
\begin{picture}(120,120)(-40,0)
\thicklines
\put( 80, 40){\circle*{5}}
\put( 64, 14){$\left(\begin{array}{cccc} 1 & 0 & 0 & 0 \\ 0 & 1 & 0 & 0 \\ 0 & 0 & 1 & 0 \\ 0 & 0 & 0 & 1 \end{array}\right)$}
\put( 80, 85){\oval(48,90)}
\put( 80,130){\circle*{5}}
\put( 64,104){$\left(\begin{array}{cccc} 0 & 0 & 1 & 0 \\ 0 & 0 & 0 & 1 \\ 1 & 0 & 0 & 0 \\ 0 & 1 & 0 & 0 \end{array}\right)$}
\put( 80, 85){\oval(90,90)}
\put(125, 80){\circle*{4}}
\put( 35, 80){\circle*{4}}
\put(132, 80){$\left(\begin{array}{cccc} 0 & 0 & 0 & 1 \\ 1 & 0 & 0 & 0 \\ 0 & 1 & 0 & 0 \\ 0 & 0 & 1 & 0 \end{array}\right)$}
\put( -5, 80){$\left(\begin{array}{cccc} 0 & 1 & 0 & 0 \\ 0 & 0 & 1 & 0 \\ 0 & 0 & 0 & 1 \\ 1 & 0 & 0 & 0\end{array}\right)$}
\end{picture}
\caption{One maximal 4-cycle and one non-maximal 2-cycle of the group P(4) of $4 \times 4$ permutation matrices.}
\label{graf2}
\end{figure}

As an example, we investigate the group P(4), representing all possible
reversible circuits with 2~bits at the input and 2~bits at the output.
There exist $4! = 24$ such circuits. 
Because P(4) is isomorphic to {\bf S}$_4$,
we investigate the cycle graph of this symmetric group.
It consists of six 2-cycles, four 3-cycles, and three 4-cycles.
No longer cycles exist, as $L(4)=4$. We have a closer look at one of the 4-cycles:
\[
q   = \left( \begin{array}{cccc} 0 & 1 & 0 & 0 \\ 0 & 0 & 1 & 0 \\ 0 & 0 & 0 & 1 \\ 1 & 0 & 0 & 0 \end{array} \right), \ \ 
q^2 = \left( \begin{array}{cccc} 0 & 0 & 1 & 0 \\ 0 & 0 & 0 & 1 \\ 1 & 0 & 0 & 0 \\ 0 & 1 & 0 & 0 \end{array} \right), \ \ 
q^3 = \left( \begin{array}{cccc} 0 & 0 & 0 & 1 \\ 1 & 0 & 0 & 0 \\ 0 & 1 & 0 & 0 \\ 0 & 0 & 1 & 0 \end{array} \right), \ \
q^4 = \left( \begin{array}{cccc} 1 & 0 & 0 & 0 \\ 0 & 1 & 0 & 0 \\ 0 & 0 & 1 & 0 \\ 0 & 0 & 0 & 1 \end{array} \right) = q^0 \ .
\]
For the interpolation between $u$, $q$, $q^2$, and $q^3$, we find
\[
\hspace{-3mm}
m(\teta) = \frac{1}{4}\ \left( \begin{array}{llll} 1 +  x + x^2 +  x^3 & 1 - ix - x^2 + ix^3 & 1 -  x + x^2 -  x^3 & 1 + ix - x^2 - ix^3 \\
                                                   1 + ix - x^2 - ix^3 & 1 +  x + x^2 +  x^3 & 1 - ix - x^2 + ix^3 & 1 -  x + x^2 -  x^3 \\
                                                   1 -  x + x^2 -  x^3 & 1 + ix - x^2 - ix^3 & 1 +  x + x^2 +  x^3 & 1 - ix - x^2 + ix^3 \\
                                                   1 - ix - x^2 + ix^3 & 1 -  x + x^2 -  x^3 & 1 + ix - x^2 - ix^3 & 1 +  x + x^2 +  x^3 \end{array} \right) \ .
\]
We note that the matrix $Q=q^2$ not only belongs to this 4-cycle, 
but also is member of a 2-cycle (although not a maximal  2-cycle):
\[
Q   = \left( \begin{array}{cccc} 0 & 0 & 1 & 0 \\ 0 & 0 & 0 & 1 \\ 1 & 0 & 0 & 0 \\ 0 & 1 & 0 & 0 \end{array} \right), \ \ 
Q^2 = \left( \begin{array}{cccc} 1 & 0 & 0 & 0 \\ 0 & 1 & 0 & 0 \\ 0 & 0 & 1 & 0 \\ 0 & 0 & 0 & 1 \end{array} \right) = Q^0 \ .
\]
For the interpolation between $u$ and $Q$, we find the Lagrange interpolation
\[
M(\teta) = \frac{1}{2}\ \left( \begin{array}{cccc} 1+x & 0 & \ 1-x\ \  & 0 \\ 0 & \ \ 1+x\  & 0 & 1-x \\ 1-x & 0 & 1+x & 0 \\ 0 & 1-x & 0 & 1+x \end{array} \right) \ ,
\]
different from $m(\teta)$. Figure \ref{graf2} displays both the 4-cycle and the 2-cycle. 

Although we have $M(0) = m(0) = u$ and $M(\pi)=m(\pi)=Q$,
we have $M(\pi/2) \neq  m(\pi/2)$. 
Indeed, $m(\pi/2)=q$ is a permutation matrix, whereas $M(\pi/2)$ is a complex unitary matrix:
\[
m(\pi/2) = \left( \begin{array}{cccc} 0 & 1 & 0 & 0 \\ 0 & 0 & 1 & 0 \\ 0 & 0 & 0 & 1 \\ 1 & 0 & 0 & 0 \end{array} \right)  
\ \mbox{ and }\ 
M(\pi/2) = \frac{1}{2}\  \left( \begin{array}{cccc} 1+i & 0 & \ 1-i\ \  & 0 \\ 0 & \ \ 1+i\  & 0 & 1-i \\ 1-i & 0 & 1+i & 0 \\ 0 & 1-i & 0 & 1+i \end{array} \right) \ .
\]
They represent the circuits 
\[
\Qcircuit @C=3mm @R=3mm {
 & \qw   & \targ     & \qw & & &              & & & & \gate{\tt V} & \qw  \\  
 & \targ & \ctrl{-1} & \qw & & & \mbox{ and } & & & & \qw          & \qw  \hspace{5mm}  \ ,
} 
\]
repectively. 
The former is the cascade of a controlled {\tt NOT} and a {\tt NOT};
the latter is a square root of {\tt NOT} (a.k.a. a~{\tt V}~gate). 
The former is a classical circuit;
the latter is a quantum   circuit.
Both circuits are square roots of the classical circuit
\[
\Qcircuit @C=3mm @R=3mm {
 & \targ & \qw \\  
 & \qw   & \qw 
} 
\]
with matrix representation $Q$.
We close this section by remarking that the above two matrix sets
$m(\teta)$ and $M(\teta)$, both 1-dimensional interpolations between the unit matrix and the matrix~$Q$,
illustrate that a unitary interpolation between two unitary matrices
is not necessarily unique.

Efficient circuit design applying  
{\tt V} gates,    controlled {\tt V} gates, 
{\tt W} gates, or controlled {\tt W} gates \cite{rahman} \cite{sasanian},
can be interpreted as using cycles of dyadic $2 \times 2$ unitary matrices \cite{vanlaer}
as matrix~$q$ of Section~2, rather than permutation matrices. 
 
\section{Conclusion}

By Lagrange interpolation, it is possible to embed a finite cyclic group
in a 1-dimensional cyclic Lie group.
By applying this technique to a cycle of an $n \times n$ permutation matrix,
one obtains a 1-dimensional subgroup of the unitary group U($n$).
In this way, we can bridge the gap between 
classical reversible computation
(represented by permutation matrices) and 
quantum computation 
(represented by unitary matrices).

\appendix

\section{Alternative expressions for the Lagrange interpolation}

Assume the numbers $\omega^0$, $\omega^1$, $\omega^2$, ..., and $\omega^{p-1}$ 
are the $p$ solutions of the equation $x^p$~$-1$~=~0. Hence:
\[
x^p-1 = \prod_k\, (x-\omega^k) \ ,
\]
such that
\begin{equation}
\prod_{k \neq j} (x-\omega^k) = \frac{x^p-1}{x-\omega^j}\ .
\label{Pi}
\end{equation}
We apply this result twice:
\begin{itemize}
\item With $x=\omega^j\, e^{i\epsilon}$ and subsequently $\epsilon \rightarrow 0$ we obtain
      \[
      \prod_{k \neq j} (\omega^j-\omega^k) =                     \lim_{\epsilon \rightarrow 0}\ \frac{\omega^{pj}e^{ip\epsilon} - 1}{\omega^je^{i\epsilon}-\omega^j} 
                                           =                     \lim_{\epsilon \rightarrow 0}\ \frac{e^{ip\epsilon}-1}{\omega^j(e^{i\epsilon}-1)} 
                                           = \frac{1}{\omega^j}\ \lim_{\epsilon \rightarrow 0}\ \frac{ip\epsilon}{i\epsilon}  = \frac{p}{\omega^j} \ ,
      \]
      such that (\ref{1}) becomes (\ref{2}). \vspace*{2mm}
\item With $x=e^{i\teta}$ we obtain
      \[
      \prod_{k \neq j} (e^{i\teta}-\omega^k) = \frac{e^{ip\teta} - 1}{e^{i\teta}-\omega^j} \ ,
      \]
      such that (\ref{2}) becomes (\ref{3}).
\end{itemize}
Eqn (\ref{4}) is obtained from eqn (\ref{2}) by computing $\omega^j\, \prod_{k \neq j} (e^{i\teta}-\omega^k)$.
We find its value as follows:
\bea
\sum_r \ \omega^{-rj} x^r
& = & \sum_r \ (\omega^{-j} x)^r                \\[1mm]
& = & \frac{(\omega^{-j}x)^p-1}{\omega^{-j}x-1} \\[1mm]
& = & \frac{\omega^{-jp}x^p-1}{\omega^{-j}x-1}  \\[1mm]
& = & \frac{x^p-1}{\omega^{-j}x-1}              \\[1mm]
& = & \omega^j \, \frac{x^p-1}{x-\omega^j}      \\[1mm]
& = & \omega^j \prod_{k \neq j} (x-\omega^k)    \ .
\eea

\section{Proof of the group structure of the Lagrange interpolation}

Assume a general operator (e.g.\ in matrix form) $\hat{a}$ of the following form 
\[
\hat{a}=\sum_{j=0}^{p-1}a_j\hat{q}^j \ ,
\]
with $a_i\in\mathbb{C}$ elements over the complex field, and $\hat{q}^p=1$.  
From now on, we will omit the operator hat notation on the $q$ operators.  
First, a straightforward rearrangement shows that  

\begin{lemmma}
Given two operators $\hat{a}$ and $\hat{b}$, then the product operator
\[
\hat{c} = \hat{a}\hat{b}
\]
has coefficients
\[
c_i = \sum_{m=0}^ia_mb_{i-m}+\sum_{m=i+1}^{p-1}a_mb_{p+i-m} \ .
\]
\end{lemmma}
Note that, in $c_i$, the indices of the coefficients of $\hat{a}$ and $\hat{b}$ 
add up to $i$ in the first summation and to $p+i$ in the second summation.
We take the summations to be strictly increasing,
such that the second term of $c_{p-1}$ vanishes by definition.

\begin{figure}[!htb]
\begin{center}
\includegraphics{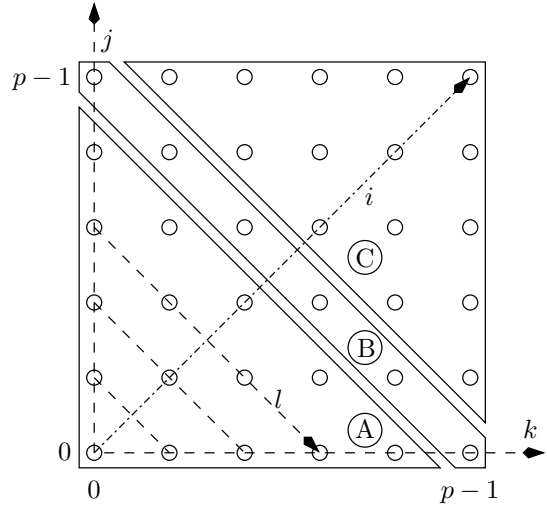}
\caption{Decomposition of the summation in eq.\ (\ref{lemma1:totalsum}).}
\label{fig:sommatie}
\end{center}
\end{figure}

{\it Proof.}
The proof is based on a simple rearrangement of terms.  
Explicitly, one can write
\begin{equation}
\hat{a}\hat{b}=\sum_{j=0}^{p-1}\sum_{k=0}^{p-1}a_jb_kq^{j+k} \ ,
\label{lemma1:totalsum}
\end{equation}
and break this double summation down in three different regions, 
corresponding to respectively regions (A), (B) and (C) in Figure~\ref{fig:sommatie},
\[
\hat{a}\hat{b} = \sum_{j=0}^{p-2}\sum_{k=0}^{p-j-2}a_jb_kq^{j+k} +
                 \sum_{j=0}^{p-1}a_jb_{p-j-1}q^{p-1} +
                 \sum_{j=1}^{p-1}\sum_{k=p-j}^{p-1}a_jb_k q^{j+k} \ .
\]

\noindent
A change of dummy summation indices $\{i=j+k,l=\frac{1}{2}(k-j)\}$ 
gives\footnote{Note that $l$ can take half-integer values.}
\bea
\hat{a}\hat{b} & = &   \sum_{i=0}^{p-2}\sum_{l=-\frac{i}{2}}^{\frac{i}{2}}a_{\frac{i}{2}-l}b_{\frac{i}{2}+l}\ q^i +
                       \sum_{l=-\frac{p}{2}+\frac{1}{2}}^{\frac{p}{2}-\frac{1}{2}}a_{\frac{p}{2}-\frac{1}{2}-l}b_{\frac{p}{2}-\frac{1}{2}+l}\ q^{p-1} \\
               &   & + \sum_{i=p}^{2p-2}\sum_{l=\frac{i}{2}-p+1}^{p-1-\frac{i}{2}}a_{\frac{i}{2}-l}b_{\frac{i}{2}+l}\ q^i \ .
\eea
The index $i$ is always larger (or equal) to $p$ in the last summation, such that 
\bea
\hat{a}\hat{b} & = &   \sum_{i=0}^{p-2}\sum_{l=-\frac{i}{2}}^{\frac{i}{2}}a_{\frac{i}{2}-l}b_{\frac{i}{2}+l}\ q^i +
                       \sum_{l=-\frac{p}{2}+\frac{1}{2}}^{\frac{p}{2}-\frac{1}{2}}a_{\frac{p}{2}-\frac{1}{2}-l}b_{\frac{p}{2}-\frac{1}{2}+l}\ q^{p-1} \\
               &   & + \sum_{m=0}^{p-2}\sum_{l=\frac{m}{2}-\frac{p}{2}+1}^{\frac{p}{2}-1-\frac{m}{2}}a_{\frac{p}{2}+\frac{m}{2}-l}b_{\frac{p}{2}+\frac{m}{2}+l}\ q^m \ ,
\eea
where $q^p=1$ has been used.  
Reintroducing dummy summation index $j=\frac{i}{2}-l$ in the first summation, 
while defining it as $j=\frac{p}{2}-\frac{1}{2}-l$ in the second, and $j=\frac{p}{2}+\frac{m}{2}-l$ in the last, 
we obtain
\[
\hat{a}\hat{b} = \sum_{i=0}^{p-2}\sum_{j=0}^i      \, a_jb_{i-j}  q^i     +
                                 \sum_{j=0}^{p-1}  \, a_jb_{p-1-j}q^{p-1} +
                 \sum_{i=0}^{p-2}\sum_{j=i+1}^{p-1}\, a_jb_{p+i-j}q^i       \ .
\]
This leads us to
\[
\hat{a}\hat{b} = \sum_{i=0}^{p-1} \left[\sum_{m=0}^ia_mb_{i-m} + \sum_{m=i+1}^{p-1}a_m b_{p+i-m}\right]q^i \ ,
\]
which completes the proof of the lemma.


The 1-parameter group structure of the Lagrange interpolation operators can now be proven. 
As demonstrated in Appendix~A, 
there are multiple representations for the $m_j$ coefficients in (\ref{m}).
We will employ here expression (\ref{3}):
\[
m_j(\teta) = \frac{1}{p}\ \frac{e^{ip\theta}-1}{e^{i\theta}-\omega^j}\ \omega^j \ .
\]

\begin{lemmma}
Given two Lagrange interpolation operators $\hat{m}(\teta_1)$ and $\hat{m}(\teta_2)$ with coefficients 
\[
m_j(\teta_k) = \frac{1}{p}\ \frac{e^{ip\theta_k}-1}{e^{i\theta_k}-\omega^j}\ \omega^j \ \mbox{ for }\ k \in \{ 1, 2\} \ ,
\]
then the product operator $\hat{a}=\hat{m}(\teta_1)\hat{m}(\teta_2)$ has coefficients
\[
a_j=\frac{1}{p}\ \frac{e^{ip(\theta_1+\theta_2)}-1}{e^{i(\theta_1+\theta_2)}-\omega^j}\ \omega^j \ .
\]
\end{lemmma}

{\it Proof.}
The previous lemma states that 
\bea
a_j & = &   \sum_{m=0}^j      \frac{1}{p}\ \frac{e^{ip\theta_1}-1}{e^{i\theta_1}-\omega^m}\ \omega^m\ \frac{1}{p}\ \frac{e^{ip\theta_2}-1}{e^{i\theta_2}-\omega^{j-m}}  \ \omega^{j-m} \\
    &   & + \sum_{m=j+1}^{p-1}\frac{1}{p}\ \frac{e^{ip\theta_1}-1}{e^{i\theta_1}-\omega^m}\ \omega^m\ \frac{1}{p}\ \frac{e^{ip\theta_2}-1}{e^{i\theta_2}-\omega^{p+j-m}}\ \omega^{p+j-m} \ .
\eea
Using $\omega^p=1$, this can be shortened to
\begin{equation}
a_j = \frac{1}{p^2}\ (e^{ip\theta_1}-1)(e^{ip\theta_2}-1)\,\omega^j
      \left[\sum_{m=0}^{p-1}\frac{1}{e^{i\theta_1}-\omega^m}\ \frac{1}{e^{i\theta_2}-\omega^{j-m}}\right] \ .
\label{1parameterproof:a12}
\end{equation}
The summation in (\ref{1parameterproof:a12}) can be reduced using partial fractions:
\begin{eqnarray}
\hspace*{-20mm}
\sum_{m=0}^{p-1} &   &           \!\!\!\frac{1}{e^{i\theta_1}-\omega^m}  \ \frac{1}{e^{i\theta_2}-\omega^{j-m}} \nonumber\\
                 & = & \frac{1}{e^{i\theta_2}}\ \sum_{m=0}^{p-1}\frac{1}{e^{i\theta_1}-\omega^m}\ \frac{\omega^m}{\omega^m-\omega^{j}e^{-i\theta_2}} \nonumber\\
                 & = & \frac{1}{e^{i(\theta_1+\theta_2)}-\omega^j}\sum_{m=0}^{p-1}\left[\frac{                          \omega^m}{e^{i\theta_1}-\omega^m}+\frac{\omega^m                                            }{\omega^m-\omega^je^{-i\theta_2}}\right] \nonumber\\
                 & = & \frac{1}{e^{i(\theta_1+\theta_2)}-\omega^j}\sum_{m=0}^{p-1}\left[\frac{e^{i\teta_1}-e^{i\teta_1}+\omega^m}{e^{i\theta_1}-\omega^m}+\frac{\omega^m-\omega^je^{-i\teta_2}+\omega^je^{-i\teta_2}}{\omega^m-\omega^je^{-i\theta_2}}\right] \nonumber\\
                 & = & \frac{1}{e^{i(\theta_1+\theta_2)}-\omega^j}\left(e^{i\theta_1}\sum_{m=0}^{p-1}\frac{1}{e^{i\theta_1}-\omega^m}-\omega^je^{-i\theta_2}\sum_{m=0}^{p-1}\frac{1}{\omega^je^{-i\theta_2}-\omega^m}\right) \, .
\label{*}
\end{eqnarray}
This formula can be even further reduced by realizing that
\begin{equation}
    \sum_{m=0}^{p-1}\prod_{k \neq m}(x-\omega^k) = \frac{d}{dx}\ \prod_k(x-\omega^k) = \frac{d}{dx}(x^p-1) = px^{p-1} 
\label{realizing}
\end{equation}
and thus
\[
\sum_{m=0}^{p-1}\frac{1}{x-\omega^m}\ =
\sum_{m=0}^{p-1}\ \frac{\prod_{k \neq m} (x-\omega^k)}{\prod_{k} (x-\omega^k)} =
\frac{1}{\prod_{k} (x-\omega^k)} \sum_{m=0}^{p-1}\ \prod_{k \neq m} (x-\omega^k) = \frac{px^{p-1}}{x^p-1} \ .
\]
Inserting this relation in both sums of (\ref{*}), we gather
\bea
\sum_{m=0}^{p-1} &   & \!\!\!\frac{1}{e^{i\theta_1}-\omega^m}\ \frac{1}{e^{i\theta_2}-\omega^{j-m}}\\
                 & = &       \frac{1}{e^{i(\theta_1+\theta_2)}-\omega^j}\ \left(        e^{ i\theta_1}\ \frac{p               e^{ i(p-1)\theta_1}}{           e^{ ip\theta_1}-1} -
                                                                                \omega^je^{-i\theta_2}\ \frac{p\omega^{(p-1)j}e^{-i(p-1)\theta_2}}{\omega^{pj}e^{-ip\theta_2}-1}\right)\\
                 & = &       \frac{p}{e^{i(\theta_1+\theta_2)}-\omega^j}\ \frac{1-e^{ip(\theta_1+\theta_2)}}{(e^{ip\theta_1}-1)(1-e^{ip\theta_2})} \ ,
\eea
in which we have used that $\omega^p=1$.  
Inserting this result in eq.\ (\ref{1parameterproof:a12}) leads to 
\[
a_j = \frac{1}{p}\ \frac{e^{ip(\theta_1+\theta_2)}-1}{e^{i(\theta_1+\theta_2)}-\omega^j}\ \omega^j 
    = m_j(\theta_1+\theta_2)\ .
\]
This means that $\hat{m}(\teta_1)\hat{m}(\teta_2)$ equals $\hat{m}(\teta_1+\teta_2)$,
which is what had to be proven.


\section{Proof of the unitarity of the Lagrange interpolation}

Because of Appendix~B we have the matrix equality $m(\teta_1)m(\teta_2)=m(\teta_1+\teta_2)$, 
and hence, in particular, $m(\teta)m(-\teta)=u$ or $m(\teta)^{-1} = m(-\teta)$.
We compute $m(-\teta)$ from (\ref{1}) 
by changing from indices $j$ and $k$ to indices $J=p-j$ and $K=p-k$:
\bea
m(-\teta) & = & \sum_j \ \frac{\prod_{k \neq j} (e^{-i\teta}-\omega^k)}           {\prod_{k \neq j} (\omega^j           -\omega^k)}           \  q^j            \\[1mm]
          & = & \sum_J \ \frac{\prod_{K \neq J} (e^{-i\teta}-\omega^{p-K})}       {\prod_{K \neq J} (\omega^{p-J}       -\omega^{p-K})}       \  q^{p-J}        \\[1mm]
          & = & \sum_J \ \frac{\prod_{K \neq J} (e^{-i\teta}-\omega^{ -K})}       {\prod_{K \neq J} (\omega^{ -J}       -\omega^{ -K})}       \  q^{ -J}        \\[1mm]
          & = & \sum_J \ \frac{\prod_{K \neq J} (e^{-i\teta}-\overline{\omega}^K)}{\prod_{K \neq J} (\overline{\omega}^J-\overline{\omega}^K)}\ (q^{\dagger})^J \\[1.5mm]
          & = & [\, m(\teta)\, ]^{\dagger}\ ,
\eea
where we took into account that $q$ is unitary: $q^{-1}=q^{\dagger}$.

We can conclude that $m(\teta)^{-1}$ equals $[\, m(\teta)\, ]^{\dagger}$ 
and thus that $m(\teta)$ is unitary.

\section{Proof of the XU property of the Lagrange interpolation}

For the classical Lagrange interpolation, the sum of the Lagrange fundamental polynomials is equal to unity,
a property known as the first Cauchy relation \cite{davis} \cite{fichtner}.
We prove here that a similar property holds for the matrix interpolation (\ref{1})-(\ref{m}):
\[
\sum_j m_j(\teta) = 1\ .
\]
Indeed, by applying eqn (\ref{2}), we find:
\bea
\sum_j m_j(\teta)
 & = &    \frac{1}{p}\ \sum_j \       \omega^j         \ \prod_{k \neq j} (x-\omega^k)  \\[1mm]
 & = & -\ \frac{1}{p}\ \sum_j \ [\ (x-\omega^j) - x \ ]\ \prod_{k \neq j} (x-\omega^k)  \\
 & = & -\ \frac{1}{p}\ \sum_j \    (x-\omega^j)        \ \prod_{k \neq j} (x-\omega^k) 
       +  \frac{x}{p}\ \sum_j \                          \prod_{k \neq j} (x-\omega^k) \ .                  
\eea
Applying (\ref{Pi}) for the former sum and (\ref{realizing}) for the latter sum, we obtain
\[
-\ (x^p-1) + x^p
\]
and thus unity.

If a matrix $q$ belongs to the group XU($n$), then all $q^j$ belong to XU($n$).
Because the row sum of a sum of matrices equals the sum of the row sums of the constituent matrices,
the row sum of $m = \sum_j m_jq^j$ equals $\sum_j m_j$ times the unit row sum of $q^j$ and thus equals $\sum_j m_j$.
Because of the above Cauchy relation, this quantity is equal to~1.
Hence, the matrix $m$ has all $n$~row sums equal to unity.
We obtain a similar result for the $n$~column sums.
Hence, the interpolation matrix~$m$ belongs to XU($n$).
In fact, the matrices $m(\teta)$ constitute a 1-dimensional subgroup 
of the $(n-1)^2$-dimensional group XU($n$).

\end{document}